\documentclass[a4paper,twoside]{article}

\usepackage{epsfig}
\usepackage{subfigure}
\usepackage{calc}
\usepackage{amssymb}
\usepackage{amstext}
\usepackage{amsmath}
\usepackage{multicol}
\usepackage{pslatex}
\usepackage{fancyhdr}
\usepackage{insticc}
\usepackage[small]{caption}
\usepackage[usenames]{color}

\newcommand{\mytitle}{\uppercase{Reliable Process for Security Policy Deployment}}

\begin{document}
\onecolumn \maketitle \normalsize \vfill

\section{\uppercase{Introduction}}

\noindent Specifying, deploying and managing access control rules for
a network architecture is one of the main tasks of a security
administrator. These rules are usually implemented by different
security devices, such as firewalls, virtual private network (VPN)
tunnels and intrusion detection systems (IDSs). The configuration of
these devices must be compatible with an established security policy.
In order to ensure this compatibility in the case of a simple network
architecture, the configuration of the security devices may be
obtained directly by translating the security requirements into
packages of specific rules for each of these devices. When the
architecture is more complex and involves several security devices,
this procedure may lead to anomalies in the configuration of these
devices and become an important source of errors exploited by
potential attackers.

These anomalies can be classified as follows. \textit{Firewall
  anomalies}, also defined in the literature as \textit{intra-} and
\textit{inter-firewall} anomalies \cite{al-shaer05}, and that refer to
those conflicts that might exist within the local set of rules of a
given firewall (\textit{intra}) or between the configuration rules of
different firewalls that match the same traffic (\textit{inter});
\textit{tunneling anomalies}, which refer to those conflicts that
might exist when both firewalls and VPN tunnels match the same
traffic; and \textit{intrusion detection system anomalies}, which
refer to those conflicts that might exist when both firewalls and IDSs
match the same traffic.

There actually exist several proposals that address the problem of
managing security policies free of anomalies. In \cite{bartal}, for
example, the authors propose a refinement mechanism based on a high
level language and that performs an automatic firewall deployment
through a refinement. However, its approach is not fully
satisfactory since it does not apply a complete separation between
the abstract security policy and the security device features and
technology. The authors in \cite{fast} propose a more complete
proposal by using the \textit{Organization Based Access Control}
(OrBAC) model \cite{orbac} as a high level policy language and an
ulterior set of compilations that derive the OrBAC specifications
into specific device configurations. Unfortunately, only firewall
management is addressed in such an approach. Other approaches, such
as \cite{al-shaer05,
  vpnconfig, esorics}, on the other hand, present audit solutions for
the analysis of more complex security setups, where not only
firewalls, but also VPN devices and IDSs, are in charge of the whole
network's security. However, the main drawback of these audit
approaches relies on their lack of knowledge about a global security
policy, which is very helpful for maintenance and troubleshooting
tasks.

In this paper, we extend the refinement approach presented in
\cite{fast}, and propose a more complete refinement process to
derive not only firewall configurations, but also VPN/IPSec
(Internet Security Protocol Suite) and scenario-based IDSs
configurations. We propose a 2-steps process to (1) formally specify
the global set of security requirements by using an expressive
access control model based on OrBAC; and (2) a set of ulterior
compilations to automatically transform such an abstract security
policy into the specific configuration of each security device
deployed over the system (e.g., firewalls, VPN/IPSec tunnels and
IDSs). This strategy not only simplifies the administrator's job,
but also guarantees that the management of policies at both high and
specific level is completely free of anomalies, i.e., ambiguities,
redundancies or unnecessary details.

The rest of this paper is structured as follows.
Section~\ref{sec:relatedWork} presents some related works.
Sections~\ref{sec:secpolicyexpression} and \ref{sec:modelingTarget}
overview our strategy and introduce the main aspects of our expressive
access control model. Section~\ref{sec:algorithms} presents our
deployment algorithms. Finally, Section~\ref{sec:conclusion} closes
the paper with some conclusions and work in progress not covered in
this paper.\\

\section{\uppercase{Related Work}}
\label{sec:relatedWork}

\noindent There exist in the literature several proposals to manage
and deploy access control policies on security devices free of
anomalies. We overview in this section those works that we consider
close to ours.

A first approach presented in \cite{bartal} proposes a refinement
mechanism based on a high level language that allows administrators
to perform automatic firewall deployments. It uses the concept of
roles to define network capabilities, and propose the use of an
inheritance mechanism through a hierarchy of entities to
automatically generate permissions. However, this approach is not
fully satisfactory since it does not apply a complete separation
between the abstract security policy and the security device
features and technology. More specifically, it does not fix clear
semantics, and its concept of role becomes ambiguous. A similar
refinement approach is also presented in \cite{hassan2003}. However,
and although the authors claim that their work is based on the RBAC
model \cite{rbac}, it also presents a lack of semantics --- it seems
that they only keep from the RBAC model the concept of role. Indeed,
the specification of network entities, roles, and permission
assignments are not rigorous and does not seem to fit any reality.

\textcolor[rgb]{0.00,0.00,0.00}{The authors in \cite{Abou2} present an
  intrusion detection approach to enforce a security policy}. They
propose the use of a "neutral language" to define a global policy
which is further deployed into a heterogeneous system. As their work
is focused on a Linux protection language, the rules that cannot be
translated into file access rules are to be translated into IDS or
firewall rules. However, although the distribution of the global
policy into the system is done in a manual fashion on different
hosts/nodes, there is no algorithm explaining the choice of these
hosts/nodes that optimally respond to global security requirements.
Although some verification processes try to guarantee anomaly-free
policies, only local configurations are considered. Some drawbacks
when managing those anomalies are moreover pointed out in \cite{Blanc}
and no solution has been yet presented. Furthermore, no IPSec devices
are taken into account.

The work presented in \cite{fast} successfully applies a set of
refinement transformation to derive from an abstract security policy
based on the OrBAC model \cite{orbac} into the network's firewalls
that might be enforced. However, the network administrator has to
assist the deployment of the access control rules by indicating
which firewall implements a specific rule. For instance, concerning
a given rule stating a certain traffic is allowed (e.g., the
\textit{ftp} service) between two hosts, the administrator has to
indicate which firewalls should implement an accept rule to fulfill
this requirement. We extend in this paper this later approach by
introducing new security devices (VPN/IPSec-based tunnels and IDSs),
improving some of the previous limitations, and guaranteeing that
neither VPN nor IDS anomalies may apply over resulting setup.

Some other approaches propose to directly analyze existing
configurations in order to warn and fix inconsistencies. The work
presented in \cite{vpnconfig}, for example, concerns the analysis of
VPN overlapping tunnels in order to detect \textit{tunneling
  anomalies}. In their approach, if an access rule concerning a
protected traffic between two points is implemented by configuring
more than one IPSec overlapping tunnels, the risk is that in some
network zones the IP packets circulate without any protection. The
authors in \cite{vpnconfig} present a discovery process to detect
such situations and propose a high-level language to deal with VPN
policies. However, a significant aspect is ignored in their
approach: the whole security policy cannot be seen as two
independent aspects --- VPN tunnels and the firewall issues. They
should not be separately modeled. Otherwise, there is a risk of
conflicts at the end of their process. The use of a single access
model, as our approach does, solves this limitation and allows us to
deal with security aspects as a whole.

In \cite{al-shaer06}, a complete taxonomy of conflicts in security
policies is presented, and two main categories are proposed: (1)
\textit{intra-policy} anomalies, which refer to those conflicts that
might exist within the local configuration of security devices; and
(2) \textit{inter-policy} anomalies, which refer to those conflicts
that might exist between the configuration rules of different security
devices that match the same traffic. The authors in \cite{al-shaer05}
propose, moreover, an audit mechanism in order to discover and warn
about these anomalies. In \cite{safecomp, esorics}, some existing
limitations in \cite{al-shaer06} are pointed out, and an alternative
set of anomalies and audit algorithms to deal with these anomalies are
proposed. However, as noted in \cite{ares}, the main
drawback of these solutions relies on the lack of knowledge
about the security policy as a whole --- from a global point of view
--- which is very useful for maintenance and troubleshooting tasks.
The managing of anomalies during our refinement process, not only
guarantees equivalent results, but also keeps with such a knowledge.

Support tools can also be used to assist administrators in their task
of configuring security devices. The Cisco Security Manager
\cite{csmanager}, for example, is designed to support the security
policy deployment on a heterogeneous network involving a large
diversity of cisco-based devices. However, we observe the following
problems when using such a tool. First, it does not offer a semantic
model rich enough to express a global security policy. Although there
is the possibility of defining variables, and thus defining access
rules involving such variables, the administrator tasks are not much
simplified. The administrator always needs a global view of the
topology in order to correctly specify each rule to network devices;
there is no automatic discovery of security devices that optimally
implement an access rule involving an IP source and a destination, as
our approach does. Furthermore, the lack of a real top-down approach
as ours (cf. Section 3.1) is partially replaced by other tools ---
e.g., conflict discovery tools that need the administrator's
assistance and that unfortunately only guarantee conflict resolution
for local configurations.

\section{\uppercase{Security Policy Expression}}
\label{sec:secpolicyexpression}

\subsection{Downward Approach}
\label{sec:approach}

\begin{figure}[!h]
\begin{center}
\includegraphics[width=0.3\textwidth]{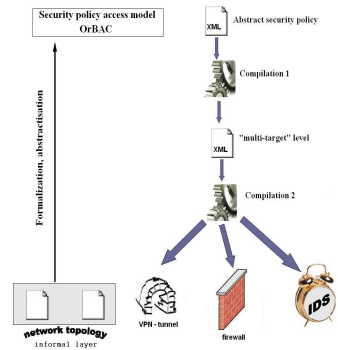}
\caption{Downward approach.\label{fig2}}
\end{center}
\end{figure}

\noindent Let us start by showing in Figure~\ref{fig2} the strategy of
our approach. The informal security requirements specified in current
language (\textit{informal layer}) are first translated into a high
level language based on the OrBAC model \cite{orbac}. Although this
translation to the OrBAC-based policy expression can not be wholly
automatic, the abstract concepts in the OrBAC model
(cf.~Section~\ref{sec:expression}) facilitate this translation for an
administrator. Based upon this abstract security policy that is
detached from any specific security device technology (e.g., NetFilter
\cite{netfilter}), we defined a set of deployment algorithms marked in
Figure~\ref{fig2}. These compilers are further detailed in Section 5.
The first compilation is iterated every time a
\textit{sub-organization} (e.g., a firewall) is revealed (in Section
3.3 we will explain the element that determines these iterations). The
result is a package of rules written in a generic expression
(\textit{multi-target}) and not for a specific technology. The second
compilation takes into account the specific technology and grammar
(syntax \& semantics) of the security devices. For example, different
transformations have to be conceived when dealing with NetFilter,
Netasq or Cisco PIX firewalls.

\subsection{Security Policy Specification}
\label{sec:expression}

\noindent OrBAC is the access control model we used to express the
abstract security policy \cite{orbac}. This model involves two
levels of abstraction: (1) an organizational level ("role",
"activity", "view" and "context" concepts); and (2) a concrete level
("subject", "action", "object") --- that are entirely compatible
with our downward approach. The OrBAC model uses first order logic
to write access control rules in the form of permissions
(\textit{Is\_permited}), prohibitions (\textit{Is\_prohibited}) and
obligations (\textit{Is\_Obliged}). For example, a permission is
derived as follows:

\medskip
\indent
\textit{$\forall$ org, $\forall$ s, $\forall$ o, $\forall$ $\alpha$, $\forall$ r, $\forall$ $\nu$, $\forall$ a, $\forall$ c\\
\indent permission(org, r, a, $\nu$, c) $\wedge$ \\
\indent empower(org, s, r) $\wedge$ use(org, o, $\nu$) $\wedge$ \\
\indent consider(org, $\alpha$, a) $\wedge$ hold(org, s, a, o, c)\\
\indent $\rightarrow$ Is\_permitted(s, $\alpha$, o)}
\medskip

If the organization \textit{org} grants role \textit{r} the permission
to perform activity \textit{a} in view \textit{$\nu$} in context
\textit{c} and if the role \textit{r} is assigned to subject
\textit{s} (\textit{empower}), the object \textit{o} is used in
\textit{$\nu$} (\textit{use}) and \textit{$\alpha$} is considered the
action implementing activity \textit{a} (\textit{consider}),
\textit{s} is granted permission to perform \textit{$\alpha$} on
\textit{o}. Let us note that the new concepts introduced by OrBAC are
the following: (1) \textit{Activity}, regrouping \textit{actions}
having common properties; (2) \textit{View}, several \textit{objects}
having the same properties on which the same rules are applied; and
(3) \textit{Context}, a concept defining the circumstances in which
some security rules can be applied.

OrBAC is based on the \textit{organization} concept assigned to each
network entity that deals with a part of the security policy. If the
(virtual) LAN the security policy is designed for, constitutes an
organization then a firewall, an IDS or an IPSec device become
\textit{sub-organizations} (organization hierarchy) of this LAN
organization. Roles are assigned to \textit{subjects}, i.e., active
entities in the network (e.g., a host, a server, a firewall
interface). A subject is assigned one or several roles and will
therefore obtain certain permissions. The notion of \textit{role}
facilitates the handling of subjects and permissions. Permissions are
obtained for each of the subjects according to their role. The
\textit{activities} are an abstraction of the network services. For
example, the action defined as "ALL\_TCP" includes all tcp network
services; "WEB" refers to https (port 443) and http (port 80).

A view regroups the \textit{objects}. As we have seen, at the
concrete level of the OrBAC model, the rules appear as
\textit{Is\_permitted(s,
  $\alpha$, o)} meaning that an entity/subject \textit{s} has the
permission to perform the action \textit{$\alpha$} on the object
\textit{o}. Hence, the object is either a network entity (e.g., a
web server) identified by its IP address or an IP packet with a
given data payload. The \textit{context} allows the definition of
specific security requirements directly at the OrBAC level. Some of
the permissions occur in a "protected" context; this leads to the
configuration of an IPSec tunnel. On the other hand, a
scenario-based IDS alert is triggered in a "vulnerability" context
associated with an attack with a known signature; a specific IP
payload may also be specified as a part of the attack signature.

Finally, OrBAC, as also RBAC does \cite{rbac}, defines role
hierarchies, and also views, activities and context hierarchies
\cite{fcs}. In the specialization/generalization hierarchy,
permissions and prohibitions are inherited \textit{downward}. These
hierarchies facilitate the administrator's task by attributing
privileges and also simplify the formalization of the security policy.

\subsection{OrBAC Security Policy}
\label{sec:secpolicy}
\begin{figure*}[!t]
\begin{center}
\includegraphics[width=0.8\textwidth]{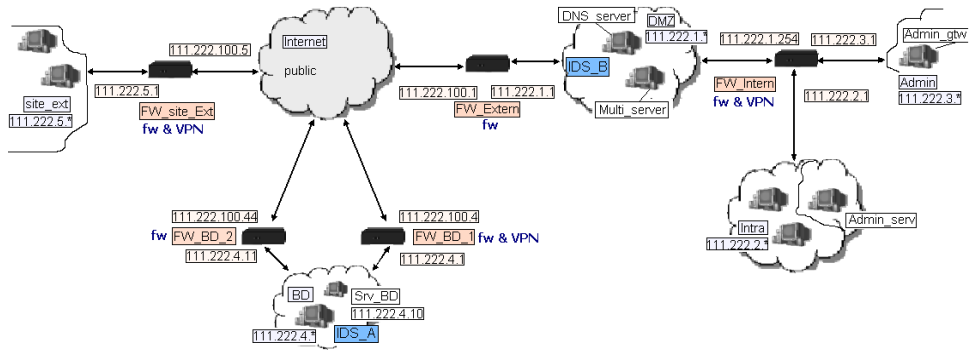}
\caption{Topology example.}\label{fig5}
\end{center}
\end{figure*}

 \noindent The authors in \cite{fast} describe
an XML-based OrBAC security policy implementation. We chose to keep
the same XML environment, but we slightly modified and proposed new
XML data structures to handle the IDS and IPSec devices. The network
architecture shown in Figure~\ref{fig5} will be used to explain our
methodology. It illustrates a private network (the "Corp" network:
111.222.0.0/16) including even geographically different sites.

Accordingly to the OrBAC model, the scheme describing the high level
policy includes an organization. It is composed of the following
parts: (1) The organization's name; (2) An element describing the
organization structure; (3) A set of rules (permissions); and, if
necessary, (4) a reference to a higher level organization
(organization hierarchy).

In the "structure" element, we distinguish the entities relevant for
the security policy (the \textit{subjects}) that compose the
network, the \textit{roles} assigned to these entities and finally,
the network services. The entities can be "host", "subnet" or
"address\_interval" types. Also, the entities exclusion is used to
simplify the structure representation. As an example, the Internet
entity is defined as 0.0.0.0/0 excluding the corporate "Corp"
network 111.222.0.0/16:

\medskip
{\footnotesize
{\setlength{\parindent}{0.5cm} \indent \verb#<#entity\verb#>#}\\
{\setlength{\parindent}{1cm} \indent \verb#<#entityName\verb#>#Net\verb#<#/entityName\verb#>#} \\
{\setlength{\parindent}{1cm} \indent \verb#<#subNet\verb#># }\\
{\setlength{\parindent}{1.5cm} \indent \verb#<#addr\verb#>#0.0.0.0\verb#<#/addr\verb#>#} \\
{\setlength{\parindent}{1.5cm} \indent \verb#<#mask\verb#>#0\verb#<#/mask\verb#>#} \\
{\setlength{\parindent}{1cm} \indent \verb#<#/subNet\verb#>#} \\
{\setlength{\parindent}{1cm} \indent \verb#<#exclusionEntity\verb#>#} \\
{\setlength{\parindent}{1.5cm} \indent \verb#<#entityName\verb#>#Corp\verb#<#/entityName\verb#>#} \\
{\setlength{\parindent}{1cm} \indent \verb#<#/exclusionEntity\verb#>#} \\
{\setlength{\parindent}{0.5cm} \indent \verb#<#/entity\verb#>#}
}
\medskip

The \textit{roles} are assigned to the entities ("entityName"). The
specialization/generalization role hierarchy is used to simplify the
OrBAC rule expression. This hierarchy is indicated by an XML child
element of the "role" element: "seniorRole". For instance, the role
"R\_FW" or "R\_VPN" is inherited by all subjects having firewall
functionalities (e.g., "FW\_Extern", in Figure 2), respectively
IPSec functionalities (e.g., "FW\_Intern"). In the following
example, the role "R\_DNS\_srv" (DNS server) is assigned to the
entity/subject "DNS\_server" that inherits the role of a server -
"R\_Srv". Thus, an access rule implying the role "R\_Srv" will
automatically be propagated to DNS server and other servers:

\medskip
{\footnotesize
{\setlength{\parindent}{0.5cm} \indent \verb#<#role\verb#>#}\\
{\setlength{\parindent}{1cm} \indent \verb#<#roleName\verb#>#R\_DNS\_srv\verb#<#/roleName\verb#>#} \\
{\setlength{\parindent}{1cm} \indent \verb#<#seniorRole\verb#># }\\
{\setlength{\parindent}{1.5cm} \indent \verb#<#roleName\verb#>#R\_Srv\verb#<#/roleName\verb#>#} \\
{\setlength{\parindent}{1cm} \indent \verb#<#/seniorRole\verb#>#} \\
{\setlength{\parindent}{1cm} \indent \verb#<#entityName\verb#>#DNS\_server\verb#<#/entityName\verb#>#} \\
{\setlength{\parindent}{0.5cm} \indent \verb#<#/role\verb#>#} }
\medskip

The permissions at the abstract level respect the data structure
shown in Figure 3. This XML schema is compliant with the OrBAC
specification. The role "roleName" performs the activity
"serviceName" on the object "target" with the role
"target/roleName". The "context" element is optional. If the
security policy does not specify any particular conditions in which
this permission is attributed to the role "roleName", the context is
"default". Otherwise, the security policy may announce a "protected"
context or a "vulnerability" one. In the first case, an IPSec-based
tunnel must be created according to the "child elements" of the
"protected" context: the type of the encryption algorithm (e.g.,
AES), the entities which have to negotiate the tunnel, a time
interval during which the IPSec tunnel is enabled, etc. The second
case will correspond to an IDS alert; a possible (XML) attribute of
the "vulnerability" context element may be the CVE vulnerability
code if known \cite{cve}. Moreover, a "content" child element of the
"context" will contain a specific data pattern as part of an attack
signature.

An important element of the "permission" is the "securityRole".
According to the OrBAC terminology, "securityRole" identifies a sub
organization. A "securityRole" is also responsible for the
activation of certain contexts, thus the activation of an access
rule. It designates the role attributed to the security device(s)
which implement(s) the corresponding access rules. For example, a
rule stating that the access from the "Internet" to the DNS server
is allowed will be implemented by the firewall "FW\_Extern"; thus,
the "securityRole" is the role "R\_FW\_Extern". Furthermore, a
permission that bounds the role "R\_Intra" and the target "Internet"
will be duplicated on both firewalls "FW\_Intern" and "FW\_Extern".
In this case, the "securityRole" regroups both "R\_FW\_Intern" and
"R\_FW\_Extern".

In the case of less complex network architectures, the
"securityRole" is given by the network administrator. Concerning
more complex architectures (and a great number of access rules),
this security role assignment is difficult to elaborate and can lead
to errors. That is why we propose some algorithms to automatically
designate the right "securityRole" under the following two
hypothesis:
\begin{figure*}
\begin{center}
\includegraphics[width=0.60\textwidth]{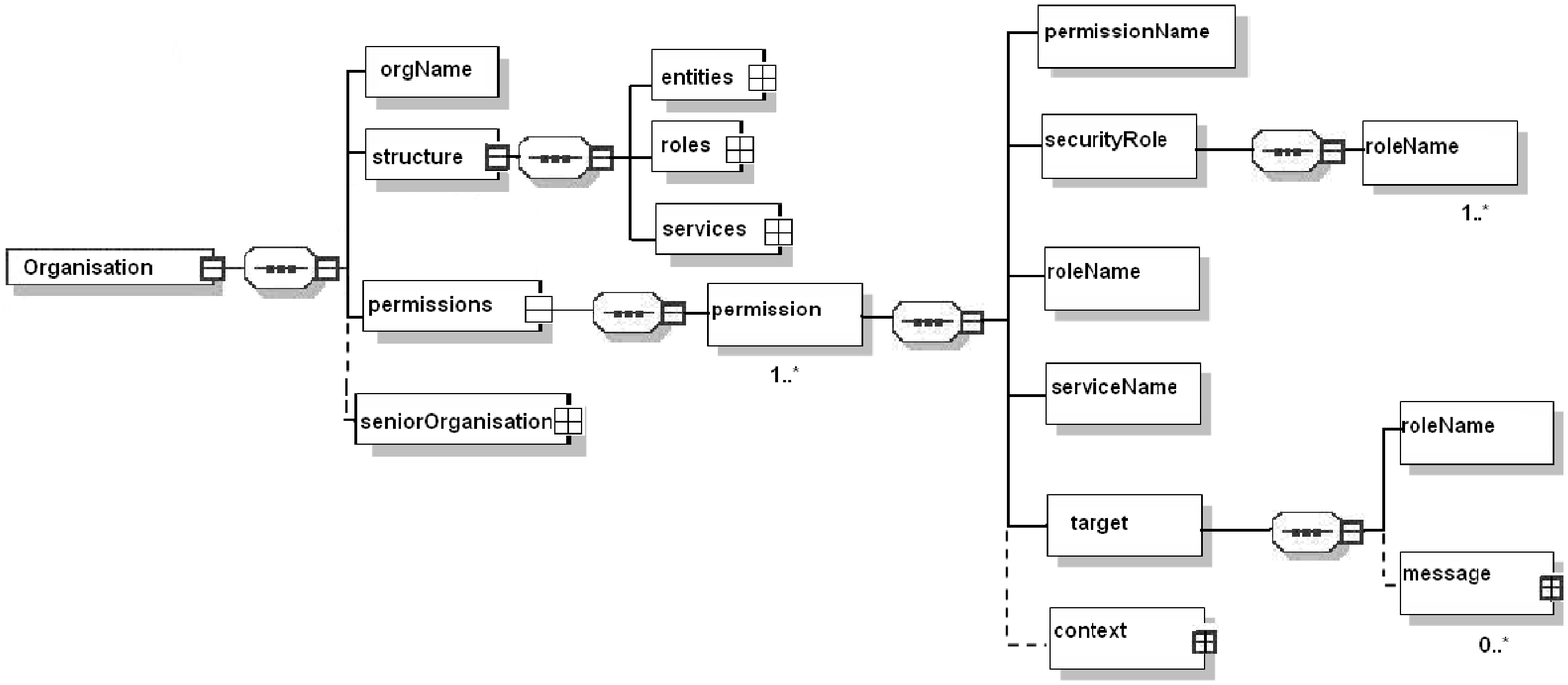}
\caption{OrBAC organization structure.} \label{fig4}
\end{center}
\end{figure*}
\begin{itemize}
\item[(1)] \textcolor[rgb]{0.00,0.00,0.00}{We consider that the formal
    policy at the OrBAC level is
    \textit{correct}\footnote{\textcolor[rgb]{0.00,0.00,0.00}{We prove
        in \cite{conflict} that such an assumption is feasible}.}}:
  the "structure" (cf.~Figure~\ref{fig4}) is well defined, without
  ambiguities (e.g., a firewall is not attributed the role of a
  server) and the access rules (cf. the "permissions" in Figure 3) are
  well specified (e.g., there is no OrBAC rule shadowed by any other).
  However, we admit that certain rules cannot be implemented because
  the adequate security device (with the appropriate functionalities)
  is missing. Our algorithms can detect this kind of mismatch when
  deploying the policy.
 \item[(2)] Inside the (virtual) network the security policy is
   designed for, the IP packets flow according to the \textit{shortest
     path} principle (as a routing protocol guarantees). The shortest
   path principle is used to identify the device(s) on which a given
   security rule must be deployed. However, notice that the shortest
   path is not always unique. For instance, in Figure 2, there are two
   possible shortest paths from site\_ext to Srv\_BD. In this case,
   our algorithms attempt to deploy the security rule on each
   candidate shortest path.
\end{itemize}

To achieve this, the OrBAC structure shown in Figure~\ref{fig4} is
parsed and relevant information about the network topology is
collected. Practically we will obtain a graph and the
\textit{shortest
  path} principle will be applied to it. We describe the methodology
in the following section.

\section{\uppercase{Modeling the Target Architecture}}
\label{sec:modelingTarget}

\subsection{Modeling the Topology}
\label{sec:modelingTopology}

\noindent At the OrBAC level, the security officer identifies the
relevant active entities (i.e., \textit{subjects}) and roles
assigned to these entities with respect to the network topology and
the security requirements. A role can be assigned to more than one
entity (e.g., all firewalls have the firewall role "R\_FW") and an
entity can have more than one role (e.g., a firewall can have IPSec
functionalities). The hierarchy of roles is defined too (e.g., a
multi-server that has the DNS\_server and Web\_server roles inherits
inevitably the server role - a less specialized role). An entity can
be either a host, a subnet or an address interval type.

As mentioned in Section~\ref{sec:secpolicy}, the entity exclusion is
used to achieve a better structuring and management of the entities.
The DMZ zone is considered to be the 111.222.1.0/24 subnet excluding
the two interfaces of the adjacent firewalls "FW\_Extern" and
"FW\_Intern" (cf.~Figure~\ref{fig5}). Multi-level exclusions may also
exist. For example, the "CorpLessIntra" is the "Corporate" entity
(111.222.0.0/16) which excludes the "Site\_ext" entity
(111.222.5.0/16) and the "Intra" entity (111.222.2.0/24) which in turn
excludes the internal interface of the firewall "FW\_Intern".
"CorpLessIntra" defines briefly all hosts unused by general corporate
employees and managed by "Admin" zone (111.222.3.0/24 - the network
administration zone).

Regarding complex network architectures, one of the most difficult
tasks for a security officer is to indicate each of the security
devices (i.e., "securityRole") which optimally implement the access
rules. In our approach, the security officer does not need to give
such an indication because our algorithms find the right set of
"securityRole" for each "permission" if "securityRole" exists. For
this purpose, the OrBAC "structure" is initially parsed and we
construct a graph where every node is a \textit{zone}\footnote{A
  \textit{zone} is either a subnet with no security device interfacing
  any other subnet or the set of interfaces of a security device}. In
order to do so, the following information is automatically extracted
during the initial phase of our process:
\begin{itemize}
\item[-] The functionalities of each security devices. For example,
  in Figure 2, the firewall "FW\_Intern" has fw \& VPN
  functionalities (firewall and IPSec capabilities).
\item[-] The set of neighbors of each zone (based on their IP
  addresses and masks\footnote{The establishing neighbors algorithm is
    based on the longest prefix matching scheme.}). For example, after
  parsing the "structure" corresponding to Figure~\ref{fig4}, the
  neighbors of the zone "FW\_Intern" are the "Intra", "DMZ" and
  "Admin" zones.
\end{itemize}

As a result of this first parsing phase, we obtain the following two
outputs:
\begin{itemize}
\item[(1)] A list of security devices, $Ss$, defined as follows:
  $Ss=\cup_j\{device_j, functionalities_j, [ \cup_n \{neighbors_{jn}
  \} ]\}$.
\item[(2)] A list of zones, $Zones$, define as follows: $Zones=\cup_i
  \{zone_i [ neighbors_{ik} \} ] \}$, and where $neighbors_{ik}$ is
  the neighbor \textit{k} zones of the \textit{i}th zone.
\end{itemize}

\subsection{Modeling Paths}
\label{sec:modelingpaths}

\noindent Let us consider a rule at the OrBAC level; it implies a
role ("roleName") and an object ("target/roleName") that corresponds
to respectively a source "Src" and a destination "Dest" entities.
This information is mandatory at the OrBAC level (cf.~Figure 3). As
already mentioned, we developed an algorithm that outputs the
optimal set of "securityRole" based on the following three
assumptions:

\begin{itemize}

\item[-] \textbf{source\_zone}: $\cup_j\{zone_j\}= Src \cap Zones$;
\item[-] \textbf{dest\_zone}: $\cup_i\{zone_i\}= Dest \cap Zones$;
\item[-] \textbf{shortest\_path} : $Zones$ \textsf{x} $Zones
  \rightarrow Ss$, such that
  shortest\_path($zone_{j}$,$zone_{i}$)$\leftarrow$$\cup_{k}
  $\verb#{#{$device_{k}$}\verb#}#.

\end{itemize}

Once identified the source and the destination zones for an access
rule, the shortest paths between a source zone and a destination
zone are computed and the security devices on this path are
revealed. Some of these security devices will be designated as
"securityRole". Moreover, the security devices on the shortest path
must have the functionalities:
\begin{itemize}
\item[-] if the access rule is in a "default" context then firewall
  functionalities are necessary;

\item[-] if the access rule is in a "protected" context then IPSec
  functionalities are required;

\item[-] if the access rule is in a "vulnerability" context, then IDS functionalities
  are required.
\end{itemize}
In a "default" context, a \textit{permission} rule will be implemented
on all firewalls found on the path; a \textit{prohibition} rule might
simply be implemented on the security device which is the closest to
the IP flow source (our \textit{shortest\_path} algorithm was designed
so as to choose the most \textit{up-stream} --- the closest to the
source --- or the most \textit{down-stream} --- the closest to the
destination --- security device). The fact that an access rule is
either a \textit{permission} or a \textit{prohibition} filtering rule
is not relevant for our discussion. That is why for didactical
reasons, we considered only \textit{permission} rules in the "default"
context.

\section{\uppercase{Deployment Algorithms}}
\label{sec:algorithms}

\noindent The main part of the security policy deployment is included
in the first compilation phase (cf. Figure~\ref{fig2}) as a set of
four processes:
\begin{itemize}
\item[-] The "structure\_parsing" with the main results, i.e., the
  list of security devices Ss\footnote{We recall, from
    Section~\ref{sec:modelingTopology}, that Ss is the list of
    security devices, such that $Ss=\cup_j\{device_j,
    functionalities_j, [ \cup_n \{neighbors_{jn} \} ]\}$.} and Zones;
\item[-] the "hierarchies\_treatement" including the role
  hierarchy treatment and the exclusion entities treatment;
\item[-] the "securityRole" phase including the shortest path computation;
\item[-] "multi-target" extracts all relevant
  information about the set of access rules the "securityRole" must implement,
  in a generic format.
\end{itemize}

\begin{figure}[!h]
\centering
\includegraphics[width=0.5\textwidth]{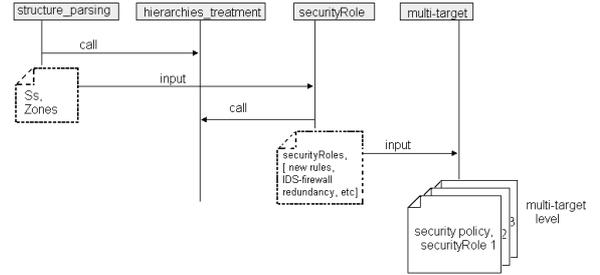}
\caption{First compilation phases.} \label{fig7}
\end{figure}

The compilation process is schematized in Figure~\ref{fig7}. The
input is the OrBAC security policy; the intermediary results are
represented by dotted lines. The final compilation result is a set
of files consisting of the part of the security policy assigned to
each "securityRole" (the "multi-target" level). "call" stands for
function callings; "input" means that the intermediary results serve
as input for other processes.

Concerning the first compilation, the main algorithms are the
"SecurityRoleDiscovery", "exclusion entities treatment" and "IDS -
Firewall\_Redundancy". From the "multi-target" level, a second
compilation is applied to obtain the packages of concrete rules
according to the specific syntax of each relevant device.

\subsection{The \textit{SecurityRoleDiscovery} Algorithm}
\label{sec:secRoleAlgorithm}

\noindent The "securityRole discovery" takes into account, as an
input, the "structure" (cf.~Figure~\ref{fig4}) parsing results:~Ss
and Zones. It also uses the \textit{shortest\_path} function. Let us
assume the "permissions" =
$\cup_{i}$\verb#{#{$permission_{i}$}\verb#}# at the OrBAC level:

\begin{figure}[!h]
\centering
\includegraphics[width=0.5\textwidth]{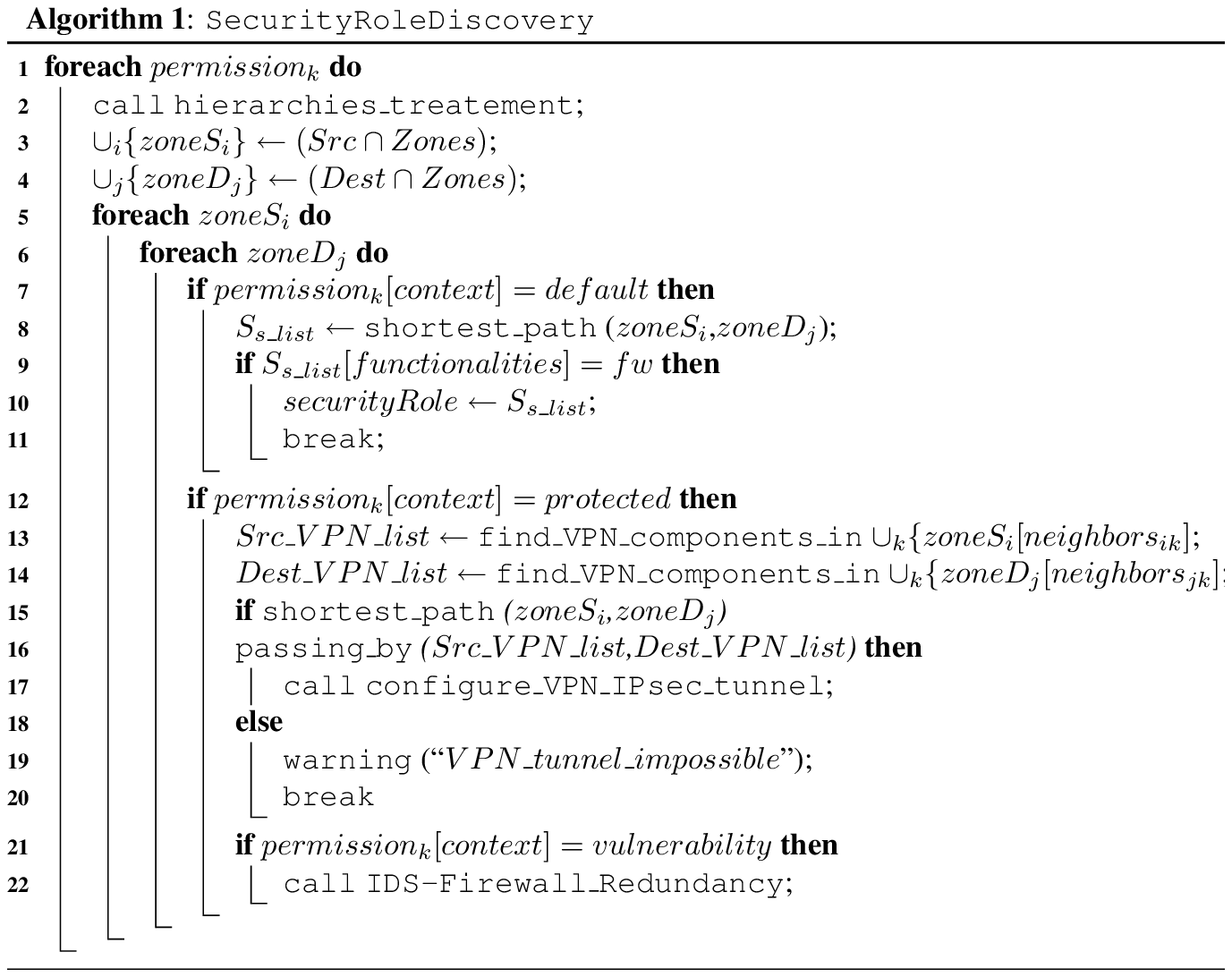}
\end{figure}

A "permission" rule in a "default" context is implemented by all
security devices with firewall functionalities on the shortest path.
For example, given the topology in
  Figure~\ref{fig5}, both firewalls "FW\_site\_Ext" and "FW\_Extern"
  implement an access rule stating that "ftp" is allowed from the
  "site\_ext" zone to the "DMZ" zone.

{\setlength{\parindent}{0.0cm}}In the "protected" context, we
formulated an extension to the shortest path function -
\textit{passing\_by()}:

{\setlength{\parindent}{0.0cm} - }it identifies the security devices
- neighbors of the source and the destination zones having IPSec
functionalities and it is supposed to compute a path between two of
them;

{\setlength{\parindent}{0.0cm} - }if a path exists, the algorithm
finds the right interfaces negotiating the IPSec tunnel (longest
prefix matching scheme);

{\setlength{\parindent}{0.0cm} - }new filtering access rules are to
be implemented on the firewalls discovered on the tunnel path (for
example, to enable the IPSec tunnel negotiation - \textit{isakmp}
with the above interfaces). This way, we avoid the conflict firewall
$\leftrightarrow$ IPSec tunnel.

Given the same topology
  (cf.~Figure 2), let us consider an access rule stating that
  all TCP traffic from the zone "Intra" to the "site\_BD" zone must be
  secured: \textit{Is\_permited(R\_Intra, ALL\_TCP,
    target-R\_site\_BD, context-protected)}. With the previous
  algorithm, the "protected" context leads to the configuration of an
  IPSec-based tunnel. The IPSec tunnel will be implemented by
  "FW\_Intern" and "FW\_BD\_1" security devices because: (1) a path
  exists from the "Intra" zone to the "site\_BD" zone and (2) the path
  crosses these two security devices with IPSec functionalities in the
  immediate neighborhood of respectively the source IP traffic zone
  ("Intra") and the destination ("site\_BD") zone. The interfaces in
  charge of the IPSec tunnel negotiation (\textit{isakmp}) are the
  "FW\_Intern" interface adjacent to the DMZ zone and respectively the
  "FW\_BD\_1" interface adjacent to the Internet zone. A filtering
  rule permitting the corresponding \textit{isakmp} traffic is
  automatically deduced and finally implemented in all firewalls on
  the tunnel path ("FW\_Intern", "FW\_Extern" and also "FW\_BD\_1").

  If the access rule is in a "vulnerability" context,
  \textit{IDS-Firewall\_Redundancy} is called
  (cf.~Section~\ref{sec:ids-fw-redundancy}). Instead of deploying an
  IDS rule, we chose to exploit an eventual IDS-firewall redundancy.
  We do not consider any IDS-IPSec tunnel interaction because IDS
  generally works on an unencrypted IP traffic.

\subsection{The \textit{exclusion entities treatment} Algorithm}
\label{sec:excTreatement}

\noindent During the first compilation, we treat the hierarchies of
roles and activities (network services). Consequently, a permission
involving the firewall role "R\_FW" will engage all entities/subjects
playing a "R\_FW" role. These entities may contain other entities
which are excluded. Deploying an access rule implying
subjects/entities which exclude other entities is solved as follows:

\begin{itemize}

\item[-] a permission at the OrBAC level involving entity E1, E1 excluding
  E2 will be translated in a generic rule which will include E1, E1
  excluding E2;

\item[-] a permission involving E1, E1 excluding E2, E2 excluding E3
  will be translated in two generic rules: the first will involve E1,
  E1 excluding E2 and the second will involve E3;

\item[-] a permission involving E1, E1 excluding E2, E2 excluding E3,
  E3 excluding E4, will be translated in two generic rules: the first
  will include E1, E1 excluding E2 and the second will include E3, E3
  excluding E4;

\item[-] a permission involving E1, E1 excluding E2 and E3, E3
  excluding E4 and E5, E5 excluding E6 will be translated in three
  generic rules: the first will include E1, E1 excluding E2 and E3,
  the second will include E4 and the third will include E5, E5
  excluding E6.

\end{itemize}

This reasoning derives from a simple mathematical logic; for the last
example, the entity E1 is defined as follows:
\begin{equation*}
E1 \wedge \overline{(E2 \vee E3) \wedge \overline{E4 \vee (E5 \wedge
\overline{E6})}} =
\end{equation*}
\begin{equation*}
 =(E1 \wedge \overline{E2 \vee E3})
\vee E4 \vee (E5 \wedge \overline{E6})
\end{equation*}

where {\setlength{\parindent}{0.0cm} "$\vee$"} stands for the addition
of a new generic permission at the multi target level
(cf.~Figure~\ref{fig7}); and "$\wedge$" denotes an exclusion.

\subsection{The \textit{IDS-Firewall\_Redundancy} Algorithm}
\label{sec:ids-fw-redundancy}

\noindent An access rule in the "vulnerability" context is deployed
on a single IDS device on the shortest path binding the source and
the destination of an IP flow. We chose the most down-stream IDS
because it is more efficient against spoofing attacks than the most
up-stream IDS. Moreover, we do not eliminate the IDS-firewall
redundancy: IDS alert sets off for IP packets that should have been
blocked by an up-stream firewall. We take advantage of this
redundancy in order to obtain relevant information regarding a
malfunctioning firewall. The "IDS-Firewall\_Redundancy" algorithm is
based on the shortest path principle with some extensions. To
illustrate our approach, let us consider the topology shown in
Figure~\ref{fig5}.

A rule in the "vulnerability" context for an IP flow with the
"Intra" source zone and the "BD" destination zone will be
implemented by IDS\_A. An analysis of the firewall configurations
located on the \textit{shortest path} connecting the "Intra" source
and the "BD" destination is launched. We choose to "analyze" only
the firewalls located up-stream; IDS\_A cannot give any information
regarding its malfunctioning down-stream firewalls.

Let us consider that the up-stream firewalls ("FW\_BD\_1",
"FW\_BD\_2", "FW\_Extern", "FW\_Intern") apply a default \textit{deny
  all} filtering policy. In this case, with our security policy
deployment, they implement only permission rules (\textit{accept} or
\textit{pass}). Each rule generally involves an IP source and
destination address and a network service (\textit{ports}). Let
(S,D,P) be the triplet including this information. The entire policy
of a firewall (e.g., "FW\_Intern") may be resumed as follows: (1)
\textbf{pass}" for F = (S$_{1}$ $\wedge$ D$_{1}$ $\wedge$ P$_{1}$)
$\vee$ (S$_{2}$ $\wedge$ D$_{2}$ $\wedge$ P$_{2}$) $\vee$ ... $\vee$
(S$_{n}$ $\wedge$ D$_{n}$ $\wedge$ P$_{n}$), where n~ =~ the filtering
rules number; (2) "\textbf{deny}" for non(F); (3) F and non(F) are
included in the "3D" space [\textit{source}, \textit{destination},
\textit{service}] where (\textit{source}, \textit{destination}) $\in$
[0.0.0.0 ; 255.255.255.255] and \textit{service} $\in$ [0 ; 65536].

We refer to the IDS-firewall redundancy only if the set of parameters
forming the firewall and the IDS rules are the same; thus,  the
(S,D,P) triplet is taken into account alongside the IDS "alert" rules
and the firewall "deny" rule. For example, there is an IDS-firewall
redundancy if an IDS alerts for the IP packets including W =
(111.222.2.0/24, 111.222.4.10, all\_tcp) (regardless of their payload)
and an up-stream firewall performs \textit{deny} for the
(111.222.2.32/27, 111.222.4.10, all\_tcp) triplet. Our
"IDS-Firewall\_Redundancy" algorithm finds the intersection between
each triplet W (corresponding to an \textit{alert}) and non($F_{i}$)
for each up-stream firewall $FW_{i}$.

 \begin{figure}[h!]
   \centering
   \includegraphics[width=0.46\textwidth]{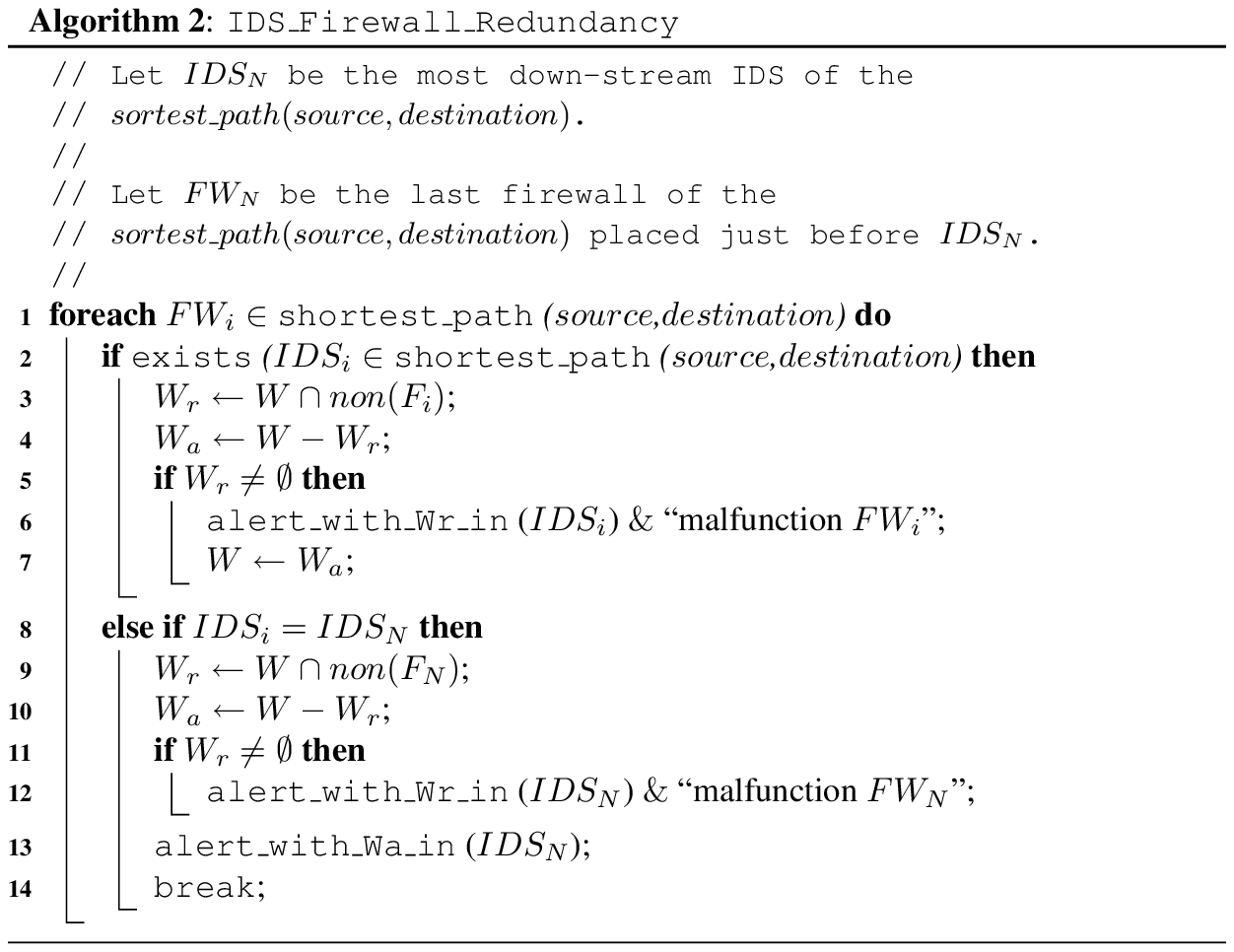}
   \label{fig10}
 \end{figure}

~~\\
Since Wr = W $\cap$ non(Fi) and Wa = W-Wr, then:
\begin{itemize}
\item[-] A first IDS rule will be implemented in the immediately
  down-stream $IDS_i$ against the firewall $FW_{i}$ the previous
  intersection was computed for. The IDS alert message will not only
  be the message corresponding to the attack but will also include
  "beware, malfunctioning $FW_{i}$". For example, in the case of
  IDS\_A$\leftrightarrow$"FW\_Intern" redundancy, involving Wr = W
  $\cap$ non(F), an IDS rule involving Wr will be implemented in
  IDS\_B and thus the "FW\_Intern" malfunction can be detected.
\item[-] The last IDS rule will be implemented in the most down-stream
  IDS and include Wa. For the previous example, an IDS rule with Wa and
  the unmodified message corresponding to the initial
  vulnerability will be implemented in IDS\_A.
\end{itemize}

\subsection{Final phase}
\label{sec:final-phase}

\noindent The second compilation phase (cf. Figure 1) translates the
generic rules to a specific security device technology. We deal with
a library of transformations. Each transformation is conditioned by
the security device features.

The intrinsic matching rule (\textit{first-matching}, or
\textit{last-matching}) and the rule order are taken into account
when designing a transformation for a firewall. Let us consider that
an entity excludes another entity in one generic
\textit{permission}; we have the choice to design either a
transformation that outputs two rules (one of them being the
\textit{negative} rule and corresponding to the excluded entity) or
a transformation in which the excluded entity is actually left out
(the result is a single \textit{pass}/\textit{accept} rule). In the
first case, the negative rule must be placed before (first-matching)
or after (last-matching) the positive rule.

NetFilter offers a mean to skip the rule order importance. The authors
in \cite{fast} use the \textit{jump} and \textit{new chain}
functionalities each time exclusion entities are involved. However,
not all firewalls we dealt with had these functionalities. The order
of rules will not be important if we succeed in writing \textit{pass}
rules and only the last rule is \textit{deny all}. We designed such a
transformation in XSLT language for Netasq F200 IPS. On the other
hand, and in order to obtain the IPSec-based tunnel configurations,
another transformation is required. Therefore, we conceived one for
the Netasq F200 family which also includes the IPSec functionalities.
As scenario-based IDS, we worked with SNORT-based IDS, for which we
considered only the alert IDS rules and a specific transformation.

\subsection{Performance evaluation}
\label{sec:evaluation}

The complete set of algorithms and processes overviewed in this
section have been implemented and evaluated in a first software
prototype. Let us briefly present in this section some of the results
we obtained. The implementation has been done by using PHP, a
general-purpose scripting language that is especially suited for web
services development. In this way, the complete refinement process can
be locally or remotely executed by using a HTTP server (e.g., Apache
server over UNIX or Windows setups) and a web browser. On the other
hand, the evaluation was carried out on an Intel-Pentium M 1.4 GHz
processor with 512 MB RAM, running Ubuntu 6.0.6 with GNU/Linux 2.6.15
(32 bits), and using Apache/2.0.55 with PHP/5.1.2 interpreter
configured.

\begin{figure}[b]
 \begin{center}
    \subfigure[Memory space evaluation.\label{fig:memoryEvaluation}]{
      \epsfig{file=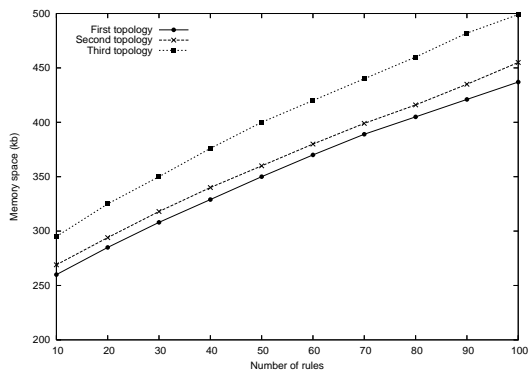, width=7cm}
    }
    \subfigure[Processing time evaluation.\label{fig:processingEvaluation}]{
      \epsfig{file=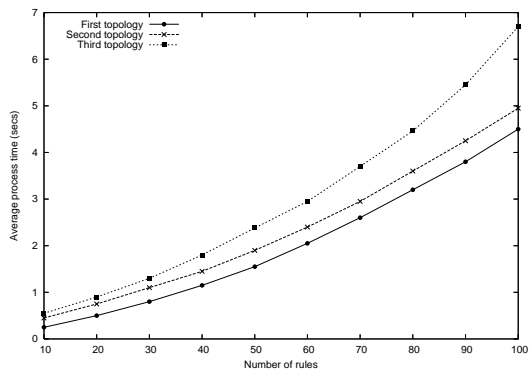, width=7cm}
    }
    \caption{Memory and processing time evaluations.}
 \end{center}
\end{figure}

For our evaluations, we specified three different IPv4 simulated
networks. The topology for the first network consisted of four
subnetworks, one SNORT-based IDS and two firewalls the access rules
were deployed on. The topology for the second network included five
subnetworks, one SNORT-based IDS, three firewalls - two of them having
IPSec capabilities. The topology for the third network consisted of
six subnetworks, two SNORT-based IDS, five firewalls - three of them
with IPSec capabilities. For each topology we considered several
security policies with an incremental number of OrBAC rules.

During the evaluation, we measured the memory space and the processing
time needed to perform the whole refinement process. The results of
these measurements are plotted in Figure~\ref{fig:memoryEvaluation}
and Figure~\ref{fig:processingEvaluation}. We can first notice in
Figure~\ref{fig:memoryEvaluation} that an important part of memory
consumption is due to the structure parsing phase (cf.
Section~\ref{sec:modelingTopology}) and then the memory increases
linearly with the OrBAC rules number. On the other hand, we notice in
Figure~\ref{fig:processingEvaluation} that the processing time is not
due to the parsing structure phase but to the OrBAC rules number and
complexity.

However, although both memory space and processing time results are
pointing out to strong requirements, we consider they are reasonable
since: (1) the implementation of our approach has been done by using a
high level scripting language, and we expect that the use of a more
efficient language will clearly improve these results; (2) our
approach relies on an off-line process which does not affect the
performance of the security policy enforcement.

We want finally to note that the implementation of our proposal in a
software prototype demonstrates the practicability of our work; and
the obtained results allow us to be very optimistic about its use in
more complex security policy scenarios.

\section{\uppercase{Conclusions}}
\label{sec:conclusion}

\noindent The configuration of security devices is a complex and
cumbersome task. A wrong configuration of those devices may lead to
weak security polices -- easily to be bypassed by unauthorized
parties. In order to help security administrators, we have presented
in this paper a refinement mechanism to properly configure and manage
the following security devices: firewalls, VPN/IPSec-based tunnels,
and scenario-based Intrusion Detection Systems (IDSs).

Our proposal allows the administrator to formally specify security
requirements by using an expressive access control model based on
OrBAC \cite{orbac}. As a result, an abstract security policy, which is
free of ambiguities, redundancies or unnecessary details, is
automatically transformed into specific security devices
configurations. This strategy not only simplifies the security
administrator's job, but also guarantees that the resulting
configuration is free of anomalies and/or inconsistencies. The
complete set of algorithms and processes presented in this paper have
been implemented in a first software prototype, and the results of a
first evaluation have been overviewed. Such implementation
demonstrates the practicability of our work and its performance
results allow us to be very optimistic about its use in more complex
security policy scenarios.

As work in progress, we are actually studying how to extend our
approach in the case where the security architecture includes IPv6
devices. More specifically, the construction of new VPN tunnels (e.g.,
IPv6-over-IPv4) for IPv6 networks must be revised, and more
investigation has to be done in order to extend the approach presented
in this paper. In parallel to this work, we are also extending our
approach to make cooperate routing and tunneling policies.

\end{document}